# Specific heat study of the $Na_{0.3}CoO_2 \cdot 1.3H_2O$ superconductor: influence of the complex chemistry

B. G. Ueland, P. Schiffer

*Department of Physics and Materials Research Institute, Pennsylvania State University, University Park, PA 16802*

R. E. Schaak, M.L. Foo, V.L. Miller, and R. J. Cava

*Department of Chemistry and Princeton Materials Institute, Princeton University, Princeton, NJ 08540*

We report results of specific heat measurements on polycrystalline samples of the layered superconductor, $Na_{0.3}CoO_2 \cdot 1.3H_2O$. The electronic contribution to the specific heat, $\gamma$, is found to be 12.2 mJ/mol-$K^2$. The feature at the superconducting transition is rather sharp, becoming broad and strongly suppressed in an applied magnetic field. The data indicate a residual normal state electronic specific heat at low temperatures, implying that there is a sizable population of normal state electrons in the samples even below $T_c$. Inhomogeneity in the Na content, to which the superconducting state is exquisitely sensitive, appears to be the most likely explanation for these results. These results further indicate that special sample handling is required for an accurate characterization of the superconducting state in this material.

The chemical, structural, and electronic analogies between the recently reported (1) 4K superconductor $Na_{0.3}CoO_2 \cdot 1.3H_2O$ and the high $T_c$ copper oxides suggests that their superconductivity may have the same underlying origin. The crystal structure of $Na_{0.3}CoO_2 \cdot 1.3H_2O$ consists of electronically active triangular $CoO_2$ layers separated by spacer layers of sodium and water (1). The triangular $CoO_2$ lattice and the speculation that it might be a Mott-Hubbard insulator has made this superconductor of particular interest to theorists (2-7). Unfortunately, the complex synthetic chemistry and chemical instability of the superconductor under ordinary laboratory conditions (8) makes the physical characterization of this superconductor relatively difficult, hence only a handful of property studies have been reported. Here we report the characterization of the specific heat of the superconductor, including the magnitude of the electronic contribution, the jump at $T_c$, and the change in the superconducting transition in an applied magnetic field. While our peak in specific heat at $T_c$ is larger than those previously reported (9,10,11), there is strong evidence for a substantial fraction of normal state electrons well below $T_c$. We attribute this primarily to small inhomogeneities in the Na content, as Na content has been shown to drastically affect the superconducting transition temperature (12).

The $Na_{0.3}CoO_2 \cdot 1.3H_2O$ samples were prepared by chemically deintercalating sodium from $Na_{0.7}CoO_2$ using bromine as an oxidizing agent. The $Na_{0.7}CoO_2$ was prepared from $Na_2CO_3$ and $Co_3O_4$ heated overnight in $O_2$ at 800°C. A 10% molar excess of $Na_2CO_3$ was added to compensate for loss due to volatilization. One-half gram of $Na_{0.7}CoO_2$ was stirred in 20 mL of a $Br_2$ solution in acetonitrile at room temperature for five days. The bromine concentration was equivalent to a molar excess of 40X relative to the amount that would theoretically be needed to remove all of the sodium from $Na_{0.7}CoO_2$ (12). The product was washed several times with acetonitrile and then water, and then dried briefly under ambient conditions. Samples were pressed into small (~10 mg, 2 mm diameter) pellets, and were not exposed to laboratory air for more than a few minutes at any time. Pellets were pressed at only moderate pressures (less than 10,000 pounds per square inch) to prevent dehydration of the superconducting phase. Powder X-ray diffraction studies indicated that the samples consisted solely of the superconducting $Na_xCoO_2 \cdot 1.3H_2O$ structure type (1,8).

It is important to note that careful preparation and handling of this material are crucial to the control of its physical properties. In particular, the superconductor is made by Na removal from the host material, $Na_{0.7}CoO_2$, in a room temperature process that is diffusion controlled. Thus, inhomogeneities in Na content may be expected if insufficient time is allowed for the full deintercalation process. This is a serious issue for the polycrystalline samples we have studied, but the problem will be orders of magnitude greater in the preparation of single crystals (10,11) where the diffusion distances are very long (mm) compared to those in polycrystalline materials (μm). In addition, the room temperature hydration-dehydration process that transforms dehydrated non-superconducting $Na_{0.3}CoO_2$ to hydrated superconducting $Na_{0.3}CoO_2 \cdot 1.3H_2O$ and back again suffers from similar kinetic issues. This process is extremely sensitive to the relative humidity of the ambient atmosphere, requiring that samples of all kinds should be exposed to open systems (i.e. laboratory air) for the shortest possible period.

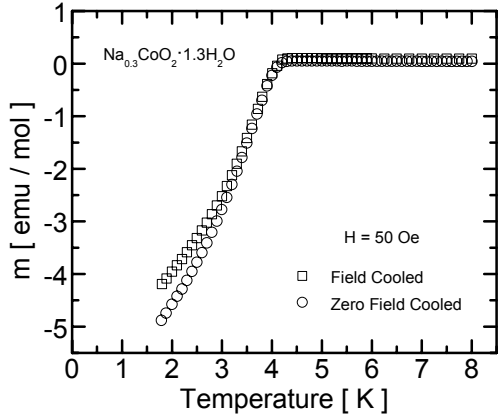

Fig. 1. The temperature dependence of the magnetization of $Na_{0.30}CoO_2 \cdot 1.3H_2O$ in a 50 Oe field.

Magnetization measurements were performed on a Quantum Design SQUID magnetometer. The specific heat was measured between temperatures of 20K and 1.8K and in magnetic fields up to 12T using the semi-adiabatic heat pulse technique of the Heat Capacity Option of a Quantum Design PPMS. Great care was taken to minimize dehydration between sample preparation and measurement. A small humidity chamber was constructed to store the samples at saturated vapor pressure subsequent to preparation. Each sample was removed from the chamber only for mounting and was exposed to ambient conditions for less than four minutes. After removal from the humidifying chamber, the weight of the pellet was monitored until the rapid weight loss associated with the loss of adsorbed water (about one minute) on exposure to lab air (8) subsided.

The sample was then immediately mounted. Following mounting, the calorimeter was inserted into the PPMS sample chamber, which was at a temperature of 274 K. The sample chamber was then purged with helium gas and cooled to 20K. An addendum measurement was performed prior to each sample measurement. Specific heat measurements were performed on two different samples to ensure reproducibility of the results.

Figure 1 shows our measurement of the temperature dependent magnetization in a 50 Oe magnetic field. As is evident from the plots, we observe a clear onset of superconductivity at $T_c \sim$ 4.2 K, consistent with results of previous studies. Figure 2 shows our raw specific heat data, plotted as $C/T$ vs. $T^2$. As can be seen from the higher temperature data in the figure, the normal state specific heat in this material can be expressed as $C(T) = \gamma T + \beta T^3$ where $\gamma$ = 12.2 mJ/molK$^2$ and $\beta$ = 3 × 10$^{-4}$ mJ/molK$^4$ between 4.2 and 9 K. These values are consistent in the different samples we have measured to within 8.6% for $\gamma$ and 55% for $\beta$ (the relatively small value of $\beta$ results in large differences between samples). Our values of $\gamma$ are comparable with the lowest of those reported elsewhere (15.9 – 14.4 mJ/mol-K$^2$ in refs 9-11).

The superconducting transition is reflected by a sharp peak in the specific heat at the same temperature as the onset of diamagnetism in $M(T)$ (i.e. $T^2$ = 16.8 K$^2$). This peak is suppressed by the application of a d.c. field, and the data taken at 12 T show only a broad peak at $T \sim$ 2.2 K. We use the high temperature specific heat as a subtraction to obtain the size of the specific heat jump in $C/T$ associated with the peak. By extrapolating with straight lines, as shown in figure 3, we obtain $\Delta C/T_c$ = 10.4 mJ/molK$^2$. This value is larger than that reported by other groups, where an anomaly in specific heat is sometimes seen well above $T_c$ ($\Delta C/T \sim$ 3-7 mJ/mol-K$^2$ at 6 K in refs 10 and 11), or at $T_c$ (~1.6 mJ/mol-K$^2$ in ref. 9).

Although our peak in $C(T)/T$ is reproducible and sharp, it does not agree quantitatively with expectations for a BCS superconductor. The specific heat jump at $T_c$ is relatively small compared to the normal state electronic specific heat. The ratio of the two is $[\Delta C/Tc]/\gamma$ = 0.85, varying by about 7% between our two samples. This ratio is considerably less than the BCS value of 1.43. While such low ratios have been observed in exotic superconductors such as $Sr_2RuO_4$ (13,14), they are highly unusual. In addition to the relatively small specific heat jump, the low temperature zero magnetic field $C/T$ data do not appear to extrapolate to zero, which should be the case if all the electrons are in the superconducting state. This could be an artifact of our data not extending to sufficiently low temperatures, but it is difficult to imagine a reasonable extrapolation which would lead to zero specific heat as $T \rightarrow 0$.

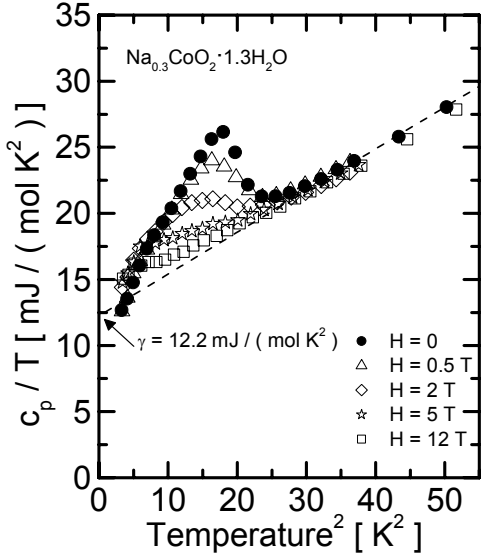

Fig. 2. The temperature dependence of the specific heat of $Na_{0.30}CoO_2 \cdot 1.3H_2O$ in various applied magnetic fields.

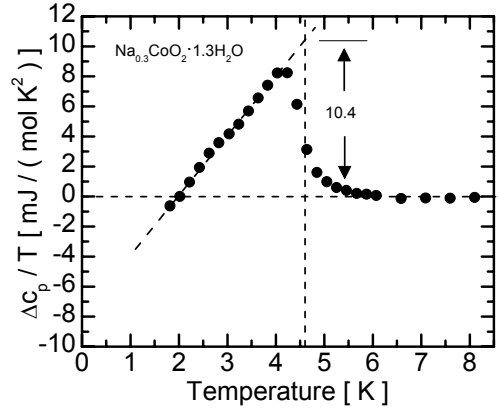

Fig. 3. The zero field specific heat peak associated with the superconducting transition after subtraction of both the phonon and electronic specific heat based on high temperature fits as described in the text.

One possible explanation for these data is that some of the measured specific heat is associated with the frozen water in our samples. The superconducting phase contains 1.3 mol $H_2O$ per formula unit, and the samples presumably also contain a small weight percentage of free water adsorbed between the grains ( < 10%). To estimate the effect on the measured specific heat, we approximate the contribution of the crystal water in $Na_{0.3}CoO_2 \cdot 1.3H_2O$ to be similar to that of pure ice. The heat capacity of ice in the 2-3 K range is 0.8 – 1.5 mJ/mol-$K^2$ (14). Including the contribution of inter-grain ice, we can assume a total of approximately 1.5 $H_2O$ per formula unit and estimate the total $H_2O$ contribution to the low temperature specific heat to be 1-2 mJ/mol-$K^2$. A substantial amount of this contribution will be accounted for in the extrapolation of the C/T vs. $T^2$ plot used to determine γ (figure 2) and in the subtraction of the $\beta T^3$ and $\gamma T$ terms in figure 3. We therefore conclude that the low temperature heat capacity of ice does not substantially influence the determination of γ, or the conclusion that there remains residual electronic specific heat below $T_c$.

Our data therefore suggest that two distinct populations of electrons exist in our material, only one of which becomes superconducting. Similar observations in early work on $Sr_2RuO_4$ were first used to support the exotic origin for the superconductivity, but improvement in sample quality eventually led to observations of a nearly zero residual specific heat at $T = 0$ (15,16). Although an unconventional microscopic mechanism cannot be ruled out as the origin for the non-BCS specific heat behavior at $T_c$, the simplest explanation is that there is a chemical inhomogeneity in our samples despite the large and relatively sharp peak in $C(T)/T$. The $C/T$ data below the peak appear to be linear in temperature rather than concave upwards, suggesting a spread in transition temperatures below 4.2 K. Based on our understanding of the complex chemistry of this material, the most likely explanation of these data is that there are slight inhomogeneities in the Na content, since $T_c$ is very sensitive to the Na concentration (12). There may also possibly be some dehydrated non-superconducting $Na_{0.3}CoO_2$ in the samples in spite of our careful preparation and handling procedures.

Our results present a challenge to the material physics community, suggesting that all samples studied to date are not fully superconducting. In addition, it is not clear why some samples show specific heat anomalies that are at 6 K, well above the $T_c$ determined by susceptibility studies, though exotic electronic behavior has been proposed as a possible explanation (11). Since the superconductivity is exquisitely sensitive to the Na content, and since the water content of the superconducting phase is unstable under ambient conditions, it will be essential to further develop preparation and characterization techniques appropriate to the unique aspects of this material before definitive physical characterization of the superconducting state will be possible.

We gratefully acknowledge support from NSF grants DMR-0244254, DMR-0101318, DMR-0213706 and DOE grant DE-FG02-98-ER45706.